\documentclass[a4paper,superscriptaddress,aps,11pt]{article}

\usepackage{graphicx,rotating,hyperref,slashed,amsmath,xcolor,amssymb,amsfonts,colortbl,cite,subcaption}
\usepackage[totalheight = 23cm, totalwidth = 17cm]{geometry}
\usepackage{physics}
\usepackage{simplewick}
\usepackage{comment}

\definecolor{bluc}{cmyk}{1,1,0,0.1}
\definecolor{rossoCP3}{cmyk}{0,.88,.77,.40}
\definecolor{rosso}{cmyk}{0,1,1,0.4}
\definecolor{rossos}{cmyk}{0,1,1,0.55}
\definecolor{rossoc}{cmyk}{0,1,1,0.2}
\definecolor{verdes}{cmyk}{0.92,0,0.59,0.4}

\hypersetup{colorlinks, bookmarksopen, bookmarksnumbered,
citecolor=verdes, linkcolor=bluc, pdfstartview=FitH, urlcolor=rossos}

\newcommand{\ud}{\,\mathrm{d}} 
\newcommand{\vect}[1]{\mathbf{#1}} 
\newcommand{\pd}[1]{\partial#1} 

\newcommand{\calA}{{\cal A}}
\newcommand{\calI}{{\cal I}}
\newcommand{\calL}{{\cal L}}

\newcommand{\calP}{{\cal P}}
\newcommand{\calR}{{\cal R}}
\newcommand{\calS}{{\cal S}}
\newcommand{\mpl}{m_{\rm Pl}}
\newcommand{\fnl}{f_{\rm NL}}

\newcommand{\mcs}{M_{\rm CS}}

\begin{document}

\begin{titlepage}

\rightline{\footnotesize{APCTP-Pre2024-016}} 
\vspace{-0.3em}
\rightline{\footnotesize{YITP-24-113}}

\begin{center}

\vskip 3em

{\LARGE \bf 
New shape for cross-bispectra in Chern-Simons gravity 
}

\vskip 3em



{\large Perseas Christodoulidis,$^{a,b}$
Jinn-Ouk Gong\footnote{jgong@ewha.ac.kr},$^{a,c}$
Wei-Chen Lin,$^{d,e}$
\vspace{0.3em}\\
Maria Mylova\footnote{maria.mylova@ipmu.jp},$^{a,f,g}$ 
and 
Misao Sasaki\footnote{misao.sasaki@ipmu.jp}$^{f,h,i}$}

\vskip 0.5cm

{\it
$^{a}$Department of Science Education, Ewha Womans University, Seoul 03760, Korea
\\
$^{b}$Korea Institute for Advanced Study, Seoul 02455, Korea
\\
$^{c}$Asia Pacific Center for Theoretical Physics, Pohang 37673, Korea
\\
$^{d}$  Extreme Physics Institute,
Pusan National University, Busan 46241, Korea
\\
$^{e}$Department of Physics, Pusan National University, Busan 46241,  Korea
\\
$^{f}$Kavli Institute for the Physics and Mathematics of the Universe (WPI)
\\
The University of Tokyo, Kashiwa, Chiba 277-8583, Japan
\\
$^{g}$Perimeter Institute for Theoretical Physics, Waterloo, Ontario N2L 2Y5, Canada
\\
$^{h}$Center for Gravitational Physics and Quantum Information
\\
Yukawa Institute for Theoretical Physics, Kyoto University, Kyoto 606-8502, Japan
\\
$^{i}$Leung Center for Cosmology and Particle Astrophysics
\\ 
National Taiwan University, Taipei 10617, Taiwan
}

\end{center}

\vskip 1.2cm

\begin{abstract}

Chern-Simons gravity is known to suffer from graviton ghost production during inflation, which suppresses the parity-violating power spectrum at scales relevant to cosmic microwave background observations. In this work, we show that allowing the initial conditions of inflation to deviate from the standard Bunch-Davies state can enhance parity-violating non-Gaussianity in the scalar-tensor cross-bispectra. Our results reveal a significant additional contribution to the cross-bispectra in the flattened configuration, offering a new avenue to constrain parity-violating gravity.





\end{abstract}

\end{titlepage}

\newpage

\tableofcontents

\section{Introduction}
\label{sec:intro}

%
%
%
%
%

\subsection{Parity violation in gravity}

Since the groundbreaking Wu experiment to test parity violation~\cite{Lee:1956qn, Wu:1957my}, substantial efforts have been devoted to understanding why parity is not a fundamental symmetry of our universe. A central question that persists is whether gravitational interactions obey the parity and time-reversal symmetries inherent to general relativity (GR).   

As GR is not a renormalizable theory, extensions of GR, driven by high energy physics considerations, are deemed necessary~\cite{donoghue1994general,Burgess:2003jk, burgess2020introduction}.
Such extensions can be addressed by constructing an effective Lagrangian that incorporates sub-leading contributions to the Einstein-Hilbert term.    
One of such is the introduction of parity-violating terms into the Einstein-Hilbert action \cite{Jackiw:2003pm, Weinberg:2008hq, Crisostomi:2017ugk, Solomon:2017nlh, Ruhdorfer:2019qmk, Alexander:2022cow}. 
This approach allows for the formulation of the most general set of operators consistent with the symmetries of the full theory, e.g. Lorentz invariance, and the fundamental degrees of freedom, such as the metric $g_{\mu\nu}$ and additional particle contents like a scalar field.  Within the framework of a low-energy expansion,  theoretical predictions remain well-controlled.

Incorporating parity-violating corrections into GR can result in asymmetric circular polarizations in the gravitational waves (GWs). That is, GWs are chiral. This can have intriguing implications  for both cosmology and astrophysics, including the early universe, which will be the focus of this work, as well as leptogenesis~\cite{Alexander:2004us, Alexander:2004xd}, compact objects~\cite{Yunes:2009hc, Alexander:2017jmt, Conroy:2019ibo, Alexander:2021ssr, Loutrel:2022tbk} and theories such as Ho\v{r}ava-Lifshitz gravity~\cite{Zhu:2013fja}.
If such an asymmetry existed in the early universe, it could further lead to potentially observable phenomena such as cosmic birefringence \cite{Alexander:2004wk, Diego-Palazuelos:2022dsq, Fujita:2022qlk}, where the polarization direction of cosmic microwave background (CMB) photons rotates during propagation. Chiral GWs could be detected by next-generation CMB experiments \cite{CMBPolStudyTeam:2008rgp, CMB-S4:2020lpa, LiteBIRD:2022cnt} or by observing a parity-violating GW background with interferometers~\cite{Bartolo:2016ami, Bartolo:2018qqn,Domcke:2019zls,Orlando:2020oko,Omiya:2023rhj}. In particular, we would expect to find detectable imprints in the CMB by observing non-vanishing TB and EB mode correlations~\cite{Lue:1998mq,  Saito:2007kt, Contaldi:2008yz, QUaD:2008ado, Gluscevic_2012}. Furthermore,  parity-violating interactions are anticipated to contribute to the higher-order correlation functions of the CMB anisotropies. Upcoming experiments aim to measure the B-mode polarization anisotropies with an expected sensitivity of $r \sim 10^{-3}$, with $r$ being tensor-to-scalar ratio~\cite{Verde:2005ff,Amblard:2006ef,CMBPolStudyTeam:2008rgp,Hazumi:2019lys}.  Confirmation of any of these signals will provide valuable insights into the nature of gravity. 

The leading-order parity-violating term allowed in a torsion-free metric theory is the gravitational Chern-Simons term $f(\phi) W \widetilde{W}$ \cite{Jackiw:2003pm, Weinberg:2008hq}, where $W_{\mu\nu\rho\sigma}$ is the Weyl tensor and $\widetilde{W}^{\mu\nu\rho\sigma} \equiv \epsilon^{\mu\nu\kappa\lambda} W_{\kappa\lambda}{}^{\rho\sigma}$, with $\epsilon^{\mu\nu\kappa\lambda}$ being the totally anti-symmetric Levi-Civita tensor. This term breaks  the parity  and time-reversal symmetries of standard Einstein gravity, while preserving  CPT symmetry as expected. Although topological in nature, this term becomes activated during inflation due to the time-dependent background.  However, the resulting chiral tensor power spectrum is severely suppressed  due to the Chern-Simons instability~\cite{Alexander:2004wk,Dyda:2012rj},  unless non-standard inflationary scenarios are considered~\cite{Satoh:2007gn,Satoh:2008ck,Satoh:2010ep,Sorbo:2011rz,Cai:2016ihp,Maleknejad:2012wqk,Mylova:2019jrj, Fujita:2023inz}.


Even in the most optimistic case, meaningful constraints for parity violation from the two-point statistics of the CMB are unlikely, as this would require a maximally chiral signal with a fairly large value of $r$~\cite{Saito:2007kt, Gluscevic_2012, Gerbino:2016mqb}. Consequently, recent researches have shifted focus to the three-point parity-violating statistics of the CMB~\cite{Maldacena:2011nz, Shiraishi:2011st, Soda:2011am, Shiraishi:2013kxa, Shiraishi:2016mok, Bartolo:2017szm, Crisostomi:2017ugk, Bartolo:2018elp,Qiao:2019hkz,Bartolo:2020gsh,Bordin:2020eui, Cabass:2021fnw, Cabass:2022jda, Philcox:2023ffy}. Interestingly, in \cite{Bartolo:2017szm} it was found that only the tensor-tensor-scalar (TTS) bispectrum is not directly suppressed by the Chern-Simons instability, being proportional to the second derivative of the coupling function $f(\phi)$  with respect to the scalar field and it could be observable by measuring the BBT or BBE angular bispectra. Nevertheless, the signal may be too weak to be detected, unless  a scenario is motivated where the Chern-Simons mass is treated as a time-dependent parameter which grows during inflation \cite{Bartolo:2018elp}.

In this study, we adopt a different approach by examining the effects on the cross-bispectra by considering initial conditions for inflation that deviate from the standard Bunch-Davies (BD) state.  Recent analysis of the parity-even cross-bispectra in general scalar-tensor gravity with non-BD initial conditions has demonstrated that the TTS non-Gaussianity (NG) can be enhanced in the flattened  (also known as folded) configuration due to the excited scalar modes, even when backreaction is taken into account \cite{Akama:2020jko}.  This enhancement can leave potentially observable imprints in the CMB.

Unlike previous studies of parity-violating cross-bispectra \cite{Bartolo:2017szm}, where the dominant NG is in the squeezed triangle configuration, in our case the dominant NG arises when the excited modes are still sub-horizon. This can potentially amplify the parity-violating cross-bispectra, resulting in a detectable signal and a new shape in the flattened configuration that has not been considered before in the literature. Recent findings have also shown that depending on the phase $\psi$ of the Bogoliubov transformation [see \eqref{eq:alpha-scalar-modes}, \eqref{eq:alpha-scalar-operators}, \eqref{eq:alpha-tensor-modes} and \eqref{eq:alpha-tensor-operators}], large tensor NG can appear in the squeezed and flattened limits of the bispectrum within Einstein gravity \cite{Kanno:2022mkx, Ghosh:2023agt}.

Prior to our analysis, it is essential to briefly discuss what we mean by non-BD  initial conditions for inflation, which is the topic of the next section.

\subsection{A short aside on the initial conditions of inflation}

An important feature of inflation is that quantum fluctuations become the seed of observable structure in the universe, with the power spectrum being almost scale-invariant.
This scale-invariant spectrum provides us with valuable information about an epoch around $50$ - $60$ $e$-folds before the end of inflationary expansion. 
However, we know little about the initial conditions of inflation. 
Any excited state could serve as an initial state just as well as the vacuum state. A straightforward and natural modification to the initial conditions of inflation involves allowing for a non-trivial initial vacuum state for a scalar field $\phi$ propagating in a de Sitter background. This approach is particularly intriguing because it enables us to test quantum field theories in curved space-time and may offer insights into the physics preceding inflation.

The usual choice of vacuum state is the adiabatic vacuum, which singles out the positive energy component in the high frequency limit, and has the property that it asymptotes to the Minkowski vacuum at very small scales. 
As a result, the stress tensor of the primordial fluctuations can be naturally regularized~\cite{birrell1984quantum}.
In the pure de Sitter space, the adiabatic vacuum corresponds to the BD state, except for the case of a minimally coupled massless scalar field for which no renormalizable de Sitter-invariant vacuum can be defined~\cite{Allen:1985ux}. 
Thus under the pure de Sitter approximation, the BD state is commonly assumed as the ground state for any subsequent cosmological evolution (for a generalisation see \cite{Ashoorioon:2018uey, Ashoorioon:2021srt}). 

The BD vacuum, however, is not a unique vacuum because it is unclear how to single out a vacuum state for quantum field theories in time-dependent space-time. 
In particular, in de Sitter space, there exists a  family of de Sitter-invariant vacua, known as the $\alpha$-vacua which are related to the BD vacuum through a Bogoliubov transformation. 
These vacua are specified by the so-called squeezing parameter $\alpha \in [0,\infty)$, which quantifies the deviation from the BD vacuum, and a phase $\psi \in (0, 2\pi)$ --  allowed by the Bogoliubov transformations \cite{Mottola:1984ar, Allen:1985ux, Higuchi:2011vw}. 
For a more geometric approach, see  \cite{Chen:2024ckx}. 
As $\alpha \rightarrow 0$, the BD vacuum is recovered, which is often used  to describe the initial state of quantum fields during the inflationary period in the early universe. 
The case $\alpha \neq 0$ has been extensively used to explore the effects of trans-Planckian fluctuations  (e.g. \cite{Martin:2000xs, Danielsson:2002kx,Danielsson:2002qh}) and to investigate  NG signatures for inflation. The latter is the main topic of this work.

Like any theory that explores the frontier of our understanding, the $\alpha$-vacuum initial condition is not without its challenges. 
Specifically,  the $\alpha$-vacua  do not satisfy the Hadamard condition, which is essential for renormalizing divergent expectation values in the stress tensor~\cite{Danielsson:2002mb, Brunetti:2005pr}. 
While the non-renormalizability of the $\alpha$-vacuum raises concerns, in practice,  the primary issue is the emergence of a divergent shape in the flattened triangle configuration,  characterized by scale-dependence proportional to, say, $(k_1+k_2-k_3)^{-n}$ for $n>0$. 
This divergence stems from the fact that in exact de Sitter space, we extend the conformal time to  $\eta \rightarrow - \infty$. One proposed solution is to limit the time that modes spend in the excited state \cite{Chen:2006nt, Holman:2007na, Gong:2013yvl, Jiang:2016nok, Mylova:2021eld}. 
While this approach may offer a remedy, a \textit{hard cutoff} may  not be universally applicable, as its effectiveness can vary depending on specific models, e.g. see the discussions in~\cite{Chen:2010xka}.
Another way to generate excited states during inflation is by introducing features which describe the transition from the BD vacuum to a non-BD one \cite{Chen:2010bka}. 
This can be seen as a \textit{soft ultraviolet completion}, offering a self-consistent mechanism to produce non-BD states, which may be preferable in the absence of a satisfactory renormalization scheme.

In general,  an effective $\alpha$-vacuum may emerge through a transition, as discussed in \cite{Holman:2007na, Chen:2010bka}, potentially driven by unknown, non-trivial quantum gravitational effects. For a transition occurring over a time scale $\Delta\eta$, the modes satisfying $k \Delta\eta \gg 1$  would remain in the BD vacuum, while those with $k \Delta\eta \ll 1$ could make a transition to the $\alpha$-vacuum. Although this transition brings the vacuum to a non-BD one, one may expect that it preserves the de Sitter symmetry at low energies. Let us therefore assume that such a transition has occurred.
The Bogoliubov coefficient $\beta$ in $\alpha$-vacuum is given by $\beta = e^{i\psi} \sinh \alpha$ with respect to the BD vacuum, where $\alpha$ and $\psi$ are constants [see \eqref{eq:alpha-scalar-modes}, \eqref{eq:alpha-scalar-operators}, \eqref{eq:alpha-tensor-modes} and \eqref{eq:alpha-tensor-operators}]. 
However, at sufficiently large momenta $k\Delta\eta \gg 1$, the state  remains in the BD state.
Thus one obtains a scale-dependent $\alpha$ such that it vanishes in the limit $k \rightarrow \infty$.

Independent of possible scenarios for realizing an $\alpha$-vacuum, we may make the ansatz $\alpha(k) \equiv \alpha_0 \exp[-(k/k_*)^n]$ ($n>0)$ to {\it regularize} the UV behavior. If we accept the transition scenario, we have $k_*\Delta\eta={\cal O}(1)$. 
If the observationally relevant $k$-modes  are much smaller than $k_*$, i.e. $k\ll k_*$,  $\alpha(k)$ becomes effectively scale-independent, and we recover the $\alpha$-vacuum properties in which the state is practically ${\rm SO}(4,1)$ invariant. This remains valid only up to the energy scale defined by the ultraviolet cutoff $k_*$. We conclude that by slightly breaking the de Sitter invariance, one can always modify the ultraviolet limit so that it satisfies the Hadamard condition.

In any case, practical limitations mean we can never achieve a perfectly flattened configuration in measurements. 
Indeed,  as it was discussed in \cite{Gong:2023kpe}, introducing a small regularization factor $k_c \ll 1$ into the CMB pipeline, represented by, say, $k_1+k_2-k_3 \to k_1 + k_2 -k_3 + k_c$ (see the discussions in Section~\ref{sec:TTS}), does not significantly alter the main characteristics of NG shapes in the flattened configuration -- this being the primary focus of our work.

Therefore, while  the non-renormalizability of the $\alpha$-vacuum  raises concerns, in practice, the breaking of de Sitter symmetries by the time-dependent background, as well the finite time of the inflationary expansion together with experimental limitations, can offer the ingredients required to device a suitable \textit{regularization} scheme.  
%
%
Additionally, the  pole structure, characteristic to theories with non-BD initial conditions, could be smoothed out by noise and dissipation effects due to other physical processes that may had taken part during inflation. Indeed, in \cite{Green:2020whw} it was argued that this \textit{singularity} could potentially provide insights into the initial conditions of inflation. This perspective gains further support from the considerations within an open effective field theory framework for inflation \cite{Burgess:2014eoa, Burgess:2015ajz, Hongo:2018ant, Salcedo:2024smn, Burgess:2024eng}.


%

In this work, we adopt a phenomenological approach and do not speculate on the exact nature of the transition from the BD vacuum to a non-BD one.  Instead, we  work directly with the $\alpha$-vacuum initial conditions which are conveniently de Sitter invariant. 
This provides a straightforward way to parameterize the potential enhancement of the bispectrum amplitude through $\alpha$ and $\psi$. For the particular purposes of this work, the absence of a specific regularization scheme is not expected to be detrimental to our results. 

Our focus is to find how parity-violating effects are enhanced in the flattened triangle limit, e.g. $k_1+k_2-k_3 \rightarrow 0$. This topic has been widely explored in the literature, with numerous works investigating scalar correlations with non-BD initial conditions \cite{Xue:2008mk, Chen:2009bc, Kundu:2011sg, Chen:2024ckx, Ansari:2024pgq, Ghosh:2024aqd, Chopping:2024oiu}. More recently, there has been a surge of interest in the graviton NG with non-BD initial conditions \cite{Kanno:2022mkx, Fumagalli:2021mpc, Gong:2023kpe,Ghosh:2023agt, Ansari:2024pgq, Akama:2024bav}. Interestingly, it was demonstrated in~\cite{Kanno:2022mkx,Gong:2023kpe} that the choice of the Bogoliubov phase $\psi$ can, in some cases, lead to significant tensor NG in the squeezed and flattened  configurations of the bispectrum in Einstein gravity.

 In this work, we further extend these considerations to address the cross-bispectra for parity-preserving and parity-violating  NG in $\alpha$-vacuum. We find that the suppression of the $\fnl$ parameter due to the Chern-Simons instability can be offset for values of the Bogoliubov phase $\psi \rightarrow \pi$ where the exponential enhancement is likely to dominate for large squeezing parameter $\alpha$, as observed in \cite{Kanno:2022mkx,Gong:2023kpe}. Additionally, sub-horizon interactions are expected to generate resonances, which can manifest as distinct signatures of NG in the flattened configuration. This presents us with a previously unexplored observational signature for parity-violating cross-bispectra.   

%
%
%
%
%

We also identify discrepancies with the scalar-scalar-tensor (SST) cross-bispectra, which were only estimated in \cite{Bartolo:2017szm}. 
Our explicit calculations show that if the mode functions evolve from the standard BD state, the SST bispectra 
vanish completely. In contrast, deviations from the typical adiabatic vacuum 
gives non-vanishing parity-violating NGs. Although the amplitude of the SST bispectra is generally suppressed to avoid excessive ghost graviton production, it can still be significantly enhanced assuming a sufficient increase in the squeezing parameter $\alpha$.

The work is organized as follows: The necessary ingredients for computing the scalar and tensor correlation functions are given in Section \ref{sec:basic}. As a warm-up we compute the cross-bispectra for standard GR in Section \ref{sec:mixed-einstein}. The physical contributions to the cross-bispectra are computed in Section \ref{sec:mixedbi-alpha} and analyzed in Section~\ref{sec:analysis}. Finally, we conclude in Section \ref{sec:conc}.









\section{Basic ingredients}
\label{sec:basic}

We start with the action for Einstein gravity, including a scalar field that serves as the inflaton, along with corrections to the Einstein-Hilbert action introduced by the gravitational Chern-Simons term:
\begin{equation}
\label{eq:action}
S = \int \dd^4 x \sqrt{-g} 
\left\{ \qty[  \frac{\mpl^2}{2}R - \frac{1}{2} g^{\mu\nu} \pd_\mu \phi \pd_\nu \phi - V(\phi) ] 
+ f(\phi) \epsilon^{\mu\nu\rho\sigma} W_{\mu\nu}{}^{\kappa\lambda} W_{\rho\sigma\kappa\lambda}
\right\}
\, ,
\end{equation}
where 
$f(\phi)$ is the dimensionless Chern-Simons coupling. Here, $\phi \equiv \varphi/M$ is a scalar field normalized with respect to some mass scale $M$, e.g. the reduced Planck mass $\mpl$. We adopt the flat gauge so that the spatial metric $h_{ij}$ is given by
\begin{equation}
h_{ij} = a^2 \big( e^\gamma \big)_{ij} 
= a^2 \bigg( \delta_{ij} + \gamma_{ij} + \frac{1}{2}\gamma_{ik}\gamma^k{}_j + \cdots \bigg)
\, ,
\end{equation}
where $\gamma^i{}_{j,i} = \gamma^i{}_i = 0$ with the spatial indices being raised and lowered by $\delta_{ij}$. These transverse and traceless conditions leave two physical degrees of freedom for $\gamma_{ij}$, which we identify as the two polarizations of tensor perturbations, or GWs. In the flat gauge, the only physical degrees of freedom are the scalar field perturbation $\delta\phi$ and $\gamma_{ij}$. To quantize these physical degrees of freedom, 
it is more convenient to expand them in terms of the Fourier modes and to use the conformal time $d\eta \equiv dt/a$. Further, we can introduce the polarization tensor $e_{ij}^s(\hat{\pmb{k}})$, with $s = +1$ and $s=-1$ denoting respectively the right- and left-circular polarizations of GWs. That is, $e_{ij}^s(\hat{\pmb{k}})$ satisfies the following properties:
\begin{align}
\label{eq:indices}
e_{ij}^s & = e_{ji}^s \, ,
\\
\label{eq:traceless}
\delta^{ij}e_{ij}^s & = 0 \, ,
\\
\label{eq:transverse}
k^ie_{ij}^s & = 0 \, ,
\\
\label{eq:2pol}
e_{ij}^s {e^{ij*s'}} & = 2 \, \delta_{ss'} \, ,
\\
\label{eq:helicity}
\frac{k_l}{k}\epsilon^{ilk}e_{jk}^s & = -ise^{i \, s}_j \, ,
\\
\label{eq:realness}
e_{ij}^s(-\hat{\pmb{k}}) & = e_{ij}^{s*}(\hat{\pmb{k}}) \, .
\end{align}
Then, the scalar and tensor perturbations are decomposed respectively as
\begin{align}
\label{eq:scalarmodeexp}
\delta\phi(\eta,\pmb{x}) 
& =
\int \frac{\ud^3k}{(2\pi)^3} e^{i\pmb{k}\cdot\pmb{x}} \delta\phi(\eta,\pmb{k}) 
\nonumber\\
& =
\int \frac{\ud^3k}{(2\pi)^3} e^{i\pmb{k}\cdot\pmb{x}} 
\Big[ a(\pmb{k}) \nu_k(\eta) + a^\dag(-\pmb{k}) \nu_k^*(\eta) \Big] 
\, ,
\\
\label{eq:tensormodeexp}
\gamma_{ij}(\eta,\pmb{x})
& =
\int \frac{\ud^3k}{(2\pi)^3} e^{i\pmb{k}\cdot\pmb{x}} 
\sum_s \gamma^s(\eta,\pmb{k}) e_{ij}^s(\hat{\pmb{k}})
\nonumber\\
& =
\int \frac{\ud^3k}{(2\pi)^3} e^{i\pmb{k}\cdot\pmb{x}} 
\sum_s \Big[ a^s(\pmb{k}) \mu_k(\eta) + a^{\dag \, s}(-\pmb{k}) \mu_k^*(\eta) \Big] e_{ij}^s(\hat{\pmb{k}})
\, ,
\end{align}
where the creation and annihilation operators obey the following canonical commutation relations:
\begin{align}
\Big[ a(\pmb{k}), a^{\dag}(\pmb{q}) \Big] & = (2\pi)^ 3\delta^{(3)}(\pmb{k}-\pmb{q}) \, ,
\\
\Big[ a^s(\pmb{k}), a^{\dag \, s'}(\pmb{q}) \Big] & = (2\pi)^3 \delta_{ss'} \delta^{(3)}(\pmb{k}-\pmb{q}) \, ,
\end{align}
otherwise zero. The BD vacuum state is defined to satisfy, for all $\pmb{k}$,
\begin{equation}
a (\pmb{k}) |0_{\pmb{k}}\rangle_\text{BD}
= a^s(\pmb{k})|0_{\pmb{k}}\rangle_\text{BD} 
= 0 \, .
\end{equation}
%
The corresponding mode functions in an exact de Sitter background, where $a = -1/(H\eta)$ with $\eta < 0$ and the Hubble parameter $H$ being constant, are given by
\begin{align}
\label{eq:scalarmode}
\nu_k(\eta) & = \frac{H}{\sqrt{2k^3}M} (1+ik\eta) e^{-ik\eta} \, ,
\\
\label{eq:tensormode}
\mu_k(\eta) & = \frac{H}{\sqrt{k^3}\mpl} (1+ik\eta)e^{-ik\eta} \, .
\end{align}
As explained in Introduction, the BD vacuum is not the only possible choice for the vacuum state. There is a family of vacuum states given by linear combinations of the BD mode functions and operators, called the $\alpha$-vacua. Therefore, we may consider another expansion of $\delta\phi$ and $\gamma_{ij}$ as in \eqref{eq:scalarmodeexp} and \eqref{eq:tensormodeexp}, but now in terms of new operators and new mode functions:
\begin{align}
\label{eq:scalarmodeexp-alpha}
\delta\phi(\eta,\pmb{x}) 
& =
\int \frac{\ud^3k}{(2\pi)^3} e^{i\pmb{k}\cdot\pmb{x}} 
\Big[ c(\pmb{k}) u_k(\eta) + c^\dag(-\pmb{k}) u_k^*(\eta) \Big] 
\, ,
\\
\label{eq:tensormodeexp-alpha}
\gamma_{ij}(\eta,\pmb{x})
& =
\int \frac{\ud^3k}{(2\pi)^3} e^{i\pmb{k}\cdot\pmb{x}} 
\sum_s \Big[ c^s(\pmb{k}) v_k(\eta) + c^{\dag \, s}(-\pmb{k}) v_k^*(\eta) \Big] e_{ij}^s(\hat{\pmb{k}})
\, ,
\end{align}
where the new operators satisfy the canonical commutation relations:
\begin{align}
\Big[ c(\pmb{k}), c^{\dag}(\pmb{q}) \Big] & = (2\pi)^ 3\delta^{(3)}(\pmb{k}-\pmb{q}) \, ,
\\
\Big[ c^s(\pmb{k}), c^{\dag \, s'}(\pmb{q}) \Big] & = (2\pi)^3 \delta_{ss'} \delta^{(3)}(\pmb{k}-\pmb{q}) \, .
\end{align}
Here, $c(\pmb{k})$ and $c^s(\pmb{k})$ annihilate the $\alpha$-vacuum:
\begin{equation}
c (\pmb{k}) |0_{\pmb{k}}\rangle_\alpha 
= c^s(\pmb{k})|0_{\pmb{k}}\rangle_\alpha 
= 0 \, .
\end{equation}
The new mode functions and the new operators of the $\alpha$-vacuum are related to those of the BD vacuum by the Bogoliubov transformation. For the scalar perturbation:
\begin{align}
\label{eq:alpha-scalar-modes}
u_k(\eta) & = \cosh\alpha \, \nu_k(\eta) + e^{i\psi}\sinh\alpha \, \nu_k^{*}(\eta) \, ,
\\
\label{eq:alpha-scalar-operators}
c(\pmb{k}) & = \cosh\alpha \, a(\pmb{k}) - e^{-i\psi}\sinh\alpha \, a^{\dag}(-\pmb{k}) \, ,
\end{align}
and similarly for tensor perturbations:
\begin{align}
\label{eq:alpha-tensor-modes}
v_k(\eta) & = \cosh\alpha \, \mu_k(\eta) + e^{i\psi}\sinh\alpha \, \mu_k^*(\eta) \, ,
\\
\label{eq:alpha-tensor-operators}
c^s(\pmb{k}) & = \cosh\alpha \, a^s(\pmb{k}) - e^{-i\psi}\sinh\alpha \, a^{\dag \, s}(-\pmb{k}) \, ,
\end{align}
where we assume $0\leq\alpha\leq\infty$ and $0\leq\psi\leq2\pi$ without losing generality.

\section{Cross-bispectra in standard Einstein gravity in $\alpha$-vacuum}
\label{sec:mixed-einstein}




As a preliminary step, we examine the cross-bispectra in standard Einstein gravity. To our knowledge, this was previously computed in the non-BD initial conditions only for the specific choice $\psi=\pi/2$~\cite{Ghosh:2023agt}, using the boostless bootstrap method. Here, we proceed our calculations using the standard in-in formalism for generic $\psi$. This comparison will be of interest when we later incorporate corrections from the Chern-Simons term.

\subsection{SST bispectrum}

%
Expanding \eqref{eq:action} up to cubic order in $\delta\phi$ and $\gamma_{ij}$, and using the relation 
\begin{equation}
\label{eq:calR}
\calR = -\frac{H}{\dot\phi_0}\delta\phi \, ,
\end{equation} 
where $\calR$ is the curvature perturbation in the comoving gauge with a dot being a derivative with respect to the cosmic time $t$, we find the following leading SST interaction\footnote{
If we have started from the beginning in the comoving gauge, we can reach \eqref{eq:cubic-sst} only after a number of partial integrations.
}:
\begin{equation} 
\label{eq:cubic-sst}
S_\text{GR}^{\rm SST} = \int \ud \eta \ud^3x  \, a^2 \epsilon\mpl^2 \mathcal{R}^{,i} \mathcal{R}^{,j} \gamma_{ij}  \, ,
\end{equation}
where $\epsilon \equiv -\dot{H}/H^2$ is the slow-roll parameter.
%
%
The corresponding interaction Hamiltonian is
\begin{align}
H_\text{GR}^{\rm SST}(\eta) 
& = 
- \int \ud^3 x \,  a^2 \epsilon \mpl^2 \mathcal{R}^{,i} \mathcal{R}^{,j}  \gamma_{ij} 
\nonumber\\
& =  
a^2 \epsilon \mpl^2 \, \int \frac{\ud^3q_1 \ud^3q_2 \ud^3q_3}{(2\pi)^{3\cdot2}}  \,  
\delta^{(3)}\left( -\pmb{q}_{123} \right)  \, 
\sum_{s_3}  q_{1}^i q_{2}^j e_{ij}^{s_3}(\hat{\pmb{q}}_3) \, 
\mathcal{R}(\pmb{q}_1)\mathcal{R} (\pmb{q}_2)  \gamma^{s_3}(\pmb{q}_3) 
\, ,
\end{align}
where $\pmb{q}_{123} \equiv \pmb{q}_1+\pmb{q}_2+\pmb{q}_3$.
%
%
The SST bispectrum is defined by
\begin{equation}
\big\langle \calR(\pmb{k}_1)\calR(\pmb{k}_2)\gamma^{s_3}(\pmb{k}_3) \big\rangle
\equiv
(2\pi)^3 \delta^{(3)}(\pmb{k}_{123}) B_\text{GR}^{s_3}(k_1,k_2,k_3)
\, ,
\end{equation}
where this correlation function is evaluated at the end of inflation, $\eta = 0$, and the vacuum expectation value is taken with respect to the $\alpha$-vacuum. 
Calculating the SST bispectrum at the end of inflation using the standard in-in formalism gives:
\begin{equation}
\big\langle \calR(0,\pmb{k}_1)\calR(0,\pmb{k}_2)\gamma^{s_3}(0,\pmb{k}_3) \big\rangle
=
2 \Im \bigg\{
\int_{-\infty}^0 \ud \eta \left\langle \calR(0,\pmb{k}_1)\calR(0,\pmb{k}_2)\gamma^{s_3}(0,\pmb{k}_3) 
H_\text{GR}^{\rm SST}(\eta) \right\rangle 
\bigg\}
\, .
\end{equation}
To proceed, we need a certain configuration of momentum vectors and the corresponding polarization tensors. Without losing generality, we can put all $\pmb{k}_i$'s lying on the $xy$-plane and align $\pmb{k}_3$ along the $x$-axis. Then we can write
\begin{align}
\pmb{k}_1 & = k_1 (\cos\phi, \sin\phi, 0) \, ,
\\
\pmb{k}_2 & = k_2 (\cos\theta, \sin\theta, 0) \, ,
\\
\pmb{k}_3 & = k_3 (1,0,0) \, ,
\end{align}
where the cosines are given by
\begin{equation}
\cos\phi = \frac{k_2^2-k_1^2-k_3^2}{2k_1k_3}
\quad \text{and} \quad
\cos\theta = \frac{k_1^2-k_2^2-k_3^2}{2k_2k_3} \, .
\end{equation}
%
%
Now, the circular polarization tensor for the tensor perturbation $\gamma_{ij}(\pmb{k}_3)$ can be written as
\begin{equation}
e_{ij}^{s_3}(\hat{\pmb{k}}_3) 
= \frac{1}{\sqrt{2}} \begin{pmatrix} 0 & 0 & 0 \\ 0 & 1 & is_3 \\ 0 & is_3 & -1 \end{pmatrix}
\, .
\end{equation}
%
Then we can find immediately, irrespective of polarizations, 
\begin{equation}
k_1^i k_2^j e_{ij}^{s_3}(\hat{\pmb{k}}_3) 
= \frac{1}{4\sqrt{2}k_3^2} \big( 2k_1^2k_2^2 + 2k_2^2k_3^2 + 2k_3^2k_1^2 - k_1^4 - k_2^4 - k_3^4 \big) \, .
\end{equation}
In addition, to keep the final result trackable, we define the relevant quantities for the various momentum combinations that appear in the final bispectrum expressions:
\begin{align}
K &\equiv \sum k_i \, , 
\\
K_{i} & \equiv K - 2 k_i \, , 
\\
P &\equiv \sum_{i<j} k_i k_j  \, , 
\\
P_{i} & \equiv P - 2 k_j k_l \quad (i \neq j \neq l \text{ and } j<l) \, .
\end{align}
Then, we find the bispectrum $B_\text{GR}^{s_3}(k_1,k_2,k_3)$ as
\begin{align}
\label{eq:Einstein-sst}
B_\text{GR}^{s_3}(k_1,k_2,k_3)
& = 
\frac{1}{4\epsilon} \frac{H^4}{\mpl^4} \frac{1}{(k_1 k_2 k_3)^2} 
\frac{k_1^ik_2^je_{ij}^{s_3}(\hat{\pmb{k}}_3)}{k_1k_2k_3}
\nonumber\\
& 
\times
\bigg\{ I_1 + \frac{1}{4} \big[ \cos(2\psi) (3I_1 + I_2) + 3 (I_1 - I_2) \big] \sinh^2(2\alpha) 
+ \frac{1}{4}(- 3I_1 + I_2)\cos\psi \sinh(4 \alpha) \bigg\}
\, ,
\end{align}
where we have defined 
\begin{align}
I_{1} & \equiv  - K + \frac{P}{K} + \frac{k_1 k_2 k_3}{K^2} \, , 
\\
I_2 &\equiv  K + \sum \frac{P_{i}}{K_{i}}  + k_1 k_2 k_3 \sum \frac{1}{K_{ i}^2} \, .
\end{align}
%
%
Note that \eqref{eq:Einstein-sst} reduces, for $\alpha=0$, to the BD result~\cite{Maldacena:2002vr}, while for $\psi = \pi/2$ to the expression given in \cite{Ghosh:2023agt}.
We also notice that \eqref{eq:Einstein-sst} is the same for both polarizations of tensor perturbations, so that parity is preserved in Einstein gravity.

\subsection{TTS bispectrum}

The leading cubic action for the TTS interactions is:
\begin{equation} 
\label{eq:cubictts}
S_\text{GR}^{\rm TTS}
= \int \ud \eta\ud^3x \frac{a^3\epsilon\mpl^2}{8} \bigg(
\calR{\gamma^{ij}}'\gamma_{ij}' + \calR\gamma^{ij,k}\gamma_{ij,k} - 2\gamma_{ij}'\gamma^{ij,k}\Delta^{-1}\calR'_{,k}
\bigg)
\, ,
\end{equation}
%
%
where a prime denotes a derivative with respect to the conformal time. The corresponding interaction Hamiltonian is
%
%
\begin{align}
H_\text{GR}^{\rm TTS}(\eta)
& =
\frac{a^2\epsilon\mpl^2}{8} \int \frac{\ud^3q_1\ud^3q_2\ud^3q_3}{(2\pi)^{3\cdot2}} \delta^{(3)}(-\pmb{q}_{123})
\sum_{s_1,s_2} e^{ij \, s_1}(\hat{\pmb{q}}_2) e_{ij}^{s_2}(\hat{\pmb{q}}_3)
\nonumber
\bigg\{ \Big[ -\gamma^{\prime s_1}(\pmb{q}_1)\gamma^{\prime s_2}(\pmb{q}_2)
\\& + (\pmb{q}_1\cdot\pmb{q}_2) \gamma^{s_1}(\pmb{q}_1)\gamma^{s_2}(\pmb{q}_2) \Big] 
\calR(\pmb{q}_3)
+ 2\frac{\pmb{q}_2\cdot\pmb{q}_3}{q_3^2} \gamma^{\prime s_1}(\pmb{q}_1)\gamma^{s_2}(\pmb{q}_2)
\calR'(\pmb{q}_3) 
\bigg\}
\, .
\end{align}
The TTS bispectrum then is calculated as
\begin{align}
\big\langle \gamma^{s_1}(0,\pmb{k}_1)\gamma^{s_2}(0,\pmb{k}_2)\calR(0,\pmb{k}_3) \big\rangle
& \equiv
(2\pi)^3 \delta^{(3)}(\pmb{k}_{123}) B_\text{GR}^{s_1 s_2}(k_1,k_2,k_3)
\nonumber\\
& =
2 \Im \bigg\{
\int_{-\infty}^0 \ud \eta \left\langle \gamma^{s_1}(0,\pmb{k}_1)\gamma^{s_2}(0,\pmb{k}_2)\calR(0,\pmb{k}_3) 
H_\text{GR}^{\rm TTS}(\eta) \right\rangle 
\bigg\}
\, .
\end{align}
%
%
For the product of the polarization tensors, we set up the three momenta in such a way that they all lie on the $xy$-plane, with the following components:
\begin{align}
\pmb{k}_1 & = k_1 (1,0,0) \, ,
\\
\pmb{k}_2 & = k_2 (\cos\theta,\sin\theta,0) \, ,
\\
\pmb{k}_3 & = k_3 (\cos\phi,\sin\phi,0) \, .
\end{align}
Then the circular polarization tensors for the tensor perturbations $\gamma_{ij}(\pmb{k}_1)$ and $\gamma_{ij}(\pmb{k}_2)$ are
\begin{align}
e_{ij}^{s_1}(\hat{\pmb{k}}_1)
& = 
\frac{1}{\sqrt{2}} \begin{pmatrix} 0 & 0 & 0 \\ 0 & 1 & is_1 \\ 0 & is_1 & -1 \end{pmatrix}
\, ,
\\
e_{ij}^{s_2}(\hat{\pmb{k}}_2)
& = 
\frac{1}{\sqrt{2}} \begin{pmatrix}
\sin^2\theta & -\sin\theta\cos\theta & -is_2\sin\theta
\\
-\sin\theta\cos\theta & \cos^2\theta & is_2\cos\theta
\\
-is_2\sin\theta & is_2\cos\theta & -1
\end{pmatrix}
\, .
\end{align}
Thus, we find
\begin{equation}
e^{ij*s_1}(\hat{\pmb{k}}_1) e_{ij}^{*s_2}(\hat{\pmb{k}}_2)
=
\frac{1}{8k_1^2k_2^2} \times
\begin{cases}
K^2K_3^2 & \text{if } ~ s_1=s_2
\vspace{0.3em}\\
K_1^2K_2^2 & \text{if } ~ s_1\neq s_2
\end{cases}
\, ,
\end{equation}
and
\begin{equation}
\pmb{k}_1\cdot\pmb{k}_2 
= \frac{k_3^2-k_1^2-k_2^2}{2} \, .
\end{equation}
Then we find the TTS bispectrum explicitly as
%
%
\begin{align}
\label{eq:GR-tts}
B_\text{GR}^{s_1s_2}(k_1,k_2,k_3)
& =
\frac{1}{8} \frac{H^4}{\mpl^4} \frac{1}{(k_1k_2k_3)^2} 
\frac{e^{ij*s_1}(\hat{\pmb{k}}_1) e_{ij}^{*s_2}(\hat{\pmb{k}}_2)}{k_1k_2k_3}
\nonumber
\bigg\{
I_3 
+ \frac{1}{2} \big[ \cos(2\psi) ( I_3 - I_4 )
\\& + 3 ( I_3 + I_4 ) \big]  \sinh^2(2\alpha)
- \bigg( I_3 + \frac{1}{2}I_4 \bigg) \cos\psi\sinh(4\alpha)
\bigg\} 
\, , 
\end{align}
where\footnote{
Note that in $I_3$ we obtained the coefficient of $k_3^3$ as $-1/2$, not $-1/4$ as in other literature. We carefully checked this coefficient a number of times by different methods and confirmed that $-1/2$ as given here is correct.
}
\begin{align}
I_3 & \equiv -{1\over 2} k_3^3 + {1\over 2} k_3 (k_1^2 + k_2^2) + {4 k_1^2 k_2^2 \over K} \, , 
\\
I_4 & \equiv 
 {16 k_1^3 k_2^3 k_3 \over K K_1 K_2 K_3   } \, .
\end{align}
%
%
Again we see that \eqref{eq:GR-tts} is the same for $s_1=s_2=\pm1$, making no distinction between left- and right-circular polarizations of tensor perturbations. Also, the cases $(s_1,s_2) = (+1,-1)$ and $(s_1,s_2) = (-1,+1)$ are identical. Thus the two independent polarizations are distinguishable but they are totally equivalent, reflecting parity conservation in GR.

\section{Cross-bispectra with Chern-Simons coupling in $\alpha$-vacuum}
\label{sec:mixedbi-alpha}


We now proceed our calculations for the case of parity-violating 
Chern-Simons term~\cite{Bartolo:2017szm}, summarized as below. Firstly, we consider the scenario where the Chern-Simons term does not introduce extra degrees of freedom into the 
standard gravity. 
That is, we do not consider the case in which the Chern-Simons term makes the transverse vector perturbation dynamical. In such a case, the vector perturbation could also interact with the scalar and tensor perturbations and complicate the discussion.
Consequently, the lapse and shift functions in the spatially flat gauge are slow-roll suppressed compared to the inflaton fluctuation, significantly simplifying the interaction Hamiltonian.
Secondly, we restrict our mode functions to the range $k_{\rm ph} < M_{\rm CS}$, where $M_{\rm CS}$ is the Chern-Simons mass scale defined by
\begin{equation}
M_{\rm CS} \equiv {\mpl^2 \over 8 \dot{f}} \, ,
\end{equation}
with $M_{\rm CS} \gg H$ to avoid ghost graviton production and to ensure that initially the physical wavelength of the modes under consideration remains well within the Hubble horizon. This condition allows us to disregard any modifications to the mode function. That is, \eqref{eq:tensormode} can be applied at the level of the three-point function. Lastly, both $H$ and $M_{\rm CS}$ are assumed to be constants during the slow-roll phase. 
With these conditions in place, we proceed to our computations.


\subsection{SST bispectrum}
\label{sec:SST}

Following \cite{Bartolo:2017szm}, the leading SST interaction is given by the Lagrangian
\begin{equation}
\calL^{\rm SST} = -8 f_\phi \sqrt{\epsilon} \delta\phi^{,l} \epsilon^{ijk} \big( \delta\phi_{,k} \gamma'_{jl,i} \big) \, ,
\end{equation}
where $f_\phi \equiv \partial f(\phi)/\partial\phi$. Going to  momentum space the corresponding interaction Hamiltonian  can be written as
\begin{equation}
H^{\rm SST} = \int \frac{\ud^3q_1\ud^3q_2\ud^3q_3}{(2\pi)^{3\cdot2}} \delta^{(3)}(-\pmb{q}_{123})
\sum_{s} \Big[ -8 f_\phi \sqrt{\epsilon} s q_1^i q_2^j q_3 e_{ij}^s(\hat{\pmb{q}}_3) \Big]
\delta\phi(\eta,\pmb{q}_1) \delta\phi(\eta,\pmb{q}_2) \gamma^{\prime s}(\eta,\pmb{q}_3)
\, .
\end{equation}
%
It is straightforward to calculate the SST cross-bispectrum, evaluated at the end of inflation, as
\begin{align}
\label{eq:2deltaphi-1tensor}
\big\langle \delta\phi(0,\pmb{k}_1) \delta\phi(0,\pmb{k}_2) \gamma_{ij}^s(0,\pmb{k}_3) \big\rangle 
\nonumber
& =
(2\pi)^3 \delta^{(3)}(\pmb{k}_{123}) \cdot 2 \Im \Bigg\{ 
-8 f_\phi \sqrt{\epsilon} \Bigg[ \sum_{s} s k_1^a k_2^b k_3 
e_{ab}^{*s}(\hat{\pmb{k}}_3) e_{ij}^s(\hat{\pmb{k}}_3) \Bigg] 
\nonumber 
\\& \times u_{k_1}(0) u_{k_2}(0) v_{k_3}(0) \int_{-\infty}^0 \ud \eta \, u_{k_1}^*(\eta) u_{k_2}^*(\eta) {v_{k_3}^*}'(\eta) 
+ (\pmb{k}_1 \leftrightarrow \pmb{k}_2) \Bigg\} \, .
\end{align}
We can  perform the integral with respect to $\eta$ 
to find
\begin{align}
\label{eq:sst-timeint}
& u_{k_1}(0) u_{k_2}(0) v_{k_3}(0) \int_{-\infty}^0 \ud \eta \, u_{k_1}^*(\eta) u_{k_2}^*(\eta) v_{k_3}^{*\prime}(\eta) 
\nonumber
 =
i \frac{H^6}{8(k_1k_2k_3)^3 \mpl^2 M^4} k_3^2 \sin\psi \sinh(2\alpha) 
\\& \times \big[ \cosh(2\alpha) - \cos\psi \sinh(2\alpha) \big]
\nonumber
\bigg\{
- 3 \bigg[ \frac{1}{K^2} + \frac{2(k_1+k_2)}{K^3} + \frac{6k_1k_2}{K^4} \bigg] 
+ \bigg[ \frac{1}{K_1^2} + \frac{2(-k_1+k_2)}{K_1^3} - \frac{6k_1k_2}{K_1^4} \bigg] 
\nonumber\\
&
+ \bigg[ \frac{1}{K_2^2} + \frac{2(k_1-k_2)}{K_2^3} - \frac{6k_1k_2}{K_2^4} \bigg] 
+ \bigg[ \frac{1}{K_3^2} + \frac{2(k_1+k_2)}{K_3^3} + \frac{6k_1k_2}{K_3^4} \bigg] 
\bigg\}
+ \text{real part} \, .
\end{align}
%
Next, we choose to work with the comoving curvature perturbation $\calR$, defined in \eqref{eq:calR}, instead of the scalar field fluctuation $\delta\phi$, 
since it is the quantity conserved during the super-horizon evolution and relevant for observational constraints.
Finally, using  the property $\gamma^s = \gamma_{ij} e^{ij*s}/2$, we obtain the SST bispectrum as 
\begin{equation}
\big\langle \calR(\pmb{k}_1) \calR(\pmb{k}_2) \gamma^{s_3}(\pmb{k}_3) \big\rangle
\equiv (2\pi)^3 \delta^{(3)}(\pmb{k}_{123}) B^{s_3}(k_1,k_2,k_3) \, ,
\end{equation}
with
\begin{align}
\label{eq:Bpvsst}
B^{s_3}(k_1, k_2, k_3)
& =
-s_3 f_\phi \sqrt{2\epsilon} \pi^4 \frac{H^2}{M^2} \frac{\calP_\calR \calP_\gamma}{(k_1k_2k_3)^2}
\frac{\sin\psi\sinh(2\alpha) \big[ \cosh(2\alpha) - \cos\psi \sinh(2\alpha) \big]}
{\big[ \cosh(2\alpha) + \cos\psi \sinh(2\alpha) \big]^2}
\nonumber\\
&
\times \frac{2k_1^2k_2^2 + 2k_2^2k_3^2 + 2k_3^2k_1^2 - k_1^4 - k_2^4 - k_3^4}{k_1k_2} \nonumber
\\& \times \bigg\{
- 3 \bigg[ \frac{1}{K^2} + \frac{2(k_1+k_2)}{K^3} + \frac{6k_1k_2}{K^4} \bigg] 
+ \bigg[ \frac{1}{K_1^2} + \frac{2(-k_1+k_2)}{K_1^3} - \frac{6k_1k_2}{K_1^4} \bigg] 
\nonumber\\
& 
+ \bigg[ \frac{1}{K_2^2} + \frac{2(k_1-k_2)}{K_2^3} - \frac{6k_1k_2}{K_2^4} \bigg] 
+ \bigg[ \frac{1}{K_3^2} + \frac{2(k_1+k_2)}{K_3^3} + \frac{6k_1k_2}{K_3^4} \bigg] 
\bigg\}
\nonumber\\
& \equiv
-s_3 f_\phi \sqrt{2\epsilon} \pi^4 \frac{H^2}{M^2} \frac{\calP_\calR \calP_\gamma}{(k_1k_2k_3)^2}
\calA_1(\alpha,\psi) \calS(k_1,k_2,k_3) \, ,
\end{align}
where we have defined the ``amplitude'' function $\calA_1$: 
\begin{equation}
\label{eq:A1}
    \calA_1(\alpha,\psi) \equiv \frac{\sin\psi\sinh(2\alpha) \big[ \cosh(2\alpha) - \cos\psi \sinh(2\alpha) \big]}
{\big[ \cosh(2\alpha) + \cos\psi \sinh(2\alpha) \big]^2} \, ,
\end{equation}
which satisfies $\calA_1(\alpha,2\pi-\psi) = -\calA_1(\alpha,\psi)$, and the dimensionless ``shape'' function $\mathcal{S}$:
\begin{align}
\calS(k_1,k_2,k_3) 
& \equiv 
\frac{2k_1^2k_2^2 + 2k_2^2k_3^2 + 2k_3^2k_1^2 - k_1^4 - k_2^4 - k_3^4}{k_1k_2} 
\nonumber\\
& 
\times
\bigg\{
- 3 \bigg[ \frac{1}{K^2} + \frac{2(k_1+k_2)}{K^3} + \frac{6k_1k_2}{K^4} \bigg] 
+ \bigg[ \frac{1}{K_1^2} + \frac{2(-k_1+k_2)}{K_1^3} - \frac{6k_1k_2}{K_1^4} \bigg] 
\nonumber\\
&   
+ \bigg[ \frac{1}{K_2^2} + \frac{2(k_1-k_2)}{K_2^3} - \frac{6k_1k_2}{K_2^4} \bigg] 
+ \bigg[ \frac{1}{K_3^2} + \frac{2(k_1+k_2)}{K_3^3} + \frac{6k_1k_2}{K_3^4} \bigg] 
\bigg\} 
\, .
\end{align}
%
%
Note that we have normalized with respect to the power spectra \cite{Kanno:2022mkx}:
%
\begin{equation}
\label{eq:powerspectrum}
\calP_\gamma = \frac{8}{\mpl^2} \bigg( \frac{H}{2\pi} \bigg)^2 
\Big[ \cosh(2\alpha) + \cos\psi \sinh(2\alpha) \Big]
= 16\epsilon \calP_\calR \, .
\end{equation}
%
%
%
It is useful to write the bispectrum in terms of $M_{\rm CS}$. Using $f_\phi = \mpl^2 / (8 M_{\rm CS} \dot\phi_0)$, we can write \eqref{eq:Bpvsst} as:
%
%
\begin{align}
\label{eq:sst-suppression}
B^{s_3}(k_1, k_2, k_3) 
& =    
-s_3 \frac{\pi^4}{8} \frac{\mpl}{M} {H \over  M_{\rm CS}}\frac{\calP_\calR \calP_\gamma}{(k_1k_2k_3)^2}
\calA_1(\alpha,\psi) \calS(k_1,k_2,k_3) \, .
\end{align}
%
%
%
%
%
%
%
%
It is thus clear that the SST bispectrum is suppressed by $H/M_{\rm CS} \ll 1$ to avoid ghost graviton production. Note that in the BD limit $\alpha\rightarrow 0$ then $\mathcal{A}_1 =0$ and the SST bispectrum \eqref{eq:Bpvsst} vanishes entirely, in contrast to the result reported in~\cite{Bartolo:2017szm}. This discrepancy may arise because their discussion of this contribution was based on an estimation, not on explicit computations. Our analysis shows that in the BD limit the SST bispectrum is purely real [see \eqref{eq:sst-timeint}] and thus vanishes. However, we find that an imaginary component, which contributes to parity violation, appears only when $\alpha \neq 0$. Therefore, 
a non-vanishing SST bispectrum is a unique signature for inflation with Chern-Simons gravity with non-BD initial states.
Moreover, as can be seen from the overall helicity factor $s_3$, the SST bispectra with left- and right-circular tensor perturbations have the opposite sign. Unlike GR, we can distinguish different polarizations so that parity is violated.

\subsection{TTS bispectrum}


We now focus on the TTS interaction. Under the slow-roll condition, the dominant interaction term is \cite{Bartolo:2017szm}
\begin{equation}
\calL^{\rm TTS} = -\epsilon^{ijk} \frac{d}{d\eta} \big( f_\phi \delta\phi \big) 
\Big( \gamma^l{}_i' \gamma_{kl,j}' - \gamma_i{}^{m,l} \gamma_{km,jl} \Big) \, .
\end{equation}
We can follow the same steps as outlined in the previous section to find the corresponding interaction Hamiltonian expressed in terms of the Fourier modes:
\begin{align}
H^{\rm TTS} 
& = 
\int \frac{\ud^3q_1\ud^3q_2\ud^3q_3}{(2\pi)^{3\cdot2}} \delta^{(3)}(-\pmb{q}_{123}) 
\sum_{s_1,s_2} 
s_2 q_2 e^{ab \, s_1}(\hat{\pmb{q}}_1) e_{ab}^{s_2}(\hat{\pmb{q}}_2) 
\nonumber\\
& 
\times 
\Big[ \gamma^{\prime s_1}(\eta,\pmb{q}_1) \gamma^{\prime s_2}(\eta,\pmb{q}_2) 
 + (\pmb{q}_1\cdot\pmb{q}_2) 
\gamma^{s_1}(\eta,\pmb{q}_1) \gamma^{s_2}(\eta,\pmb{q}_2) \Big]
\Big[ f_{\phi\phi} \phi_0' \delta\phi(\eta,\pmb{q}_3) + f_\phi \delta\phi'(\eta,\pmb{q}_3) \Big]
\, .
\end{align}
Then the resulting cross-bispectrum is written as
%
%
\begin{align}
&
\big\langle \gamma^{s_1}(0,\pmb{k}_1) \gamma^{s_2}(0,\pmb{k}_2) \delta\phi(0,\pmb{k}_3) \big\rangle
 =
(2\pi)^3 \delta^{(3)}(\pmb{k}_{123}) \cdot 2 \Im \Bigg\{
\Bigg[(s_1 k_1+s_2 k_2) e^{ab*\,s_1}(\hat{\pmb{k}}_1) e_{ab}^{*s_2}(\hat{\pmb{k}}_2)  
 \Bigg]
 \nonumber
\\& \times v_{k_1}(0) v_{k_2}(0)u_{k_3}(0)
 \bigg[ \int_{-\infty}^0 \ud \eta \, f_\phi {v_{k_1}^*}'(\eta) {v_{k_2}^*}'(\eta) {u_{k_3}^*}'(\eta)  + \int_{-\infty}^0 \ud \eta \, f_{\phi\phi} \phi_0' {v_{k_1}^*}'(\eta) {v_{k_2}^*}'(\eta) u_{k_3}^*(\eta)
\nonumber\\
&
+ (\pmb{k}_1\cdot\pmb{k}_2) \int_{-\infty}^0 \ud \eta \, f_\phi v_{k_1}^*(\eta) v_{k_2}^*(\eta) {u_{k_3}^*}'(\eta) 
+ (\pmb{k}_1\cdot\pmb{k}_2) \int_{-\infty}^0 \ud \eta \, f_{\phi\phi} \phi_0' v_{k_1}^*(\eta) v_{k_2}^*(\eta) u_{k_3}^*(\eta) \bigg] \Bigg\}
\, .
\end{align}
As before, these time integrals can be all performed analytically. We can now obtain the TTS bispectrum:
\begin{equation}
\big\langle \gamma^{s_1}(\pmb{k}_1) \gamma^{s_2}(\pmb{k}_2) \calR(\pmb{k}_3) \big\rangle
\equiv (2\pi)^3 \delta^{(3)}(\pmb{k}_{123}) B^{s_1 s_2}(k_1,k_2,k_3) \, ,
\end{equation}
with 
\begin{align}
\label{eq:TTS}
B^{s_1 s_2}(k_1,k_2,k_3)
& =
\sqrt{\frac{\epsilon}{2}} \frac{\pi^4}{2} \frac{H^2}{\mpl M} \frac{\calP_\calR \calP_\gamma}{(k_1k_2k_3)^2}  
\bigg\{
\calA_1(\alpha,\psi) \bigg( f_\phi \Big[ \calS_1(k_1,k_2,k_3) + \calS_2(k_1,k_2,k_3) \Big]
\nonumber\\
& 
+ \sqrt{\epsilon} \frac{\mpl}{M} f_{\phi\phi} \Big[ \calS_3(k_1,k_2,k_3) + \calS_4(k_1,k_2,k_3) \Big] \bigg)
+ \sqrt{\epsilon} \frac{\mpl}{M} f_{\phi\phi} \calA_2(\alpha,\psi) \calS_5(k_1,k_2,k_3) \bigg\}
\, ,
\end{align}
where $\calA_1(\alpha,\psi)$ is the same as before, given by \eqref{eq:A1}, and
%
\begin{align}
\calA_2(\alpha,\psi) 
& \equiv
\frac{3\cosh(4\alpha) - 3\cos\psi\sinh(4\alpha) - 1}{[\cosh(2\alpha) + \cos\psi\sinh(2\alpha)]^2} 
\, ,
\\
\label{eq:shapes}
\calS_i(k_1,k_2,k_3) 
& \equiv
\frac{s_1 k_1+s_2 k_2}{k_1^3k_2^3k_3} \times
\left\{
\begin{array}{ll}
K^2 K_3^2 & \text{if } ~ s_1=s_2
\vspace{0.3em}\\
K_1^2 K_2^2 & \text{if } ~ s_1\neq s_2
\end{array}
\right\}
\times \calI_i(k_1,k_2,k_3)
\, ,
\end{align}
with
\begin{align}
\label{eq:I1}
\calI_1(k_1,k_2,k_3)
& \equiv
6 \, (k_1k_2k_3)^2 \bigg( -\frac{3}{K^4} + \frac{1}{K_1^4} + \frac{1}{K_2^4} + \frac{1}{K_3^4} \bigg)
\, ,
\\
\label{eq:I2}
\calI_2(k_1,k_2,k_3)
& \equiv
\frac{k_3^2}{2} (k_1^2+k_2^2-k_3^2) \bigg\{
-3 \bigg[ 
\frac{1}{K^2} + \frac{2(k_1+k_2)}{K^3} + \frac{6k_1k_2}{K^4} \bigg]
+ \bigg[ \frac{1}{K_1^2} + \frac{2(-k_1+k_2)}{K_1^3} - \frac{6k_1k_2}{K_1^4} \bigg] 
\nonumber\\
& 
+ \bigg[ \frac{1}{K_2^2} + \frac{2(k_1-k_2)}{K_2^3} - \frac{6k_1k_2}{K_2^4} \bigg]
+ \bigg[ \frac{1}{K_3^2} + \frac{2(k_1+k_2)}{K_3^3} + \frac{6k_1k_2}{K_3^4} \bigg]
\bigg\}
\, ,
\\
\label{eq:I3}
\calI_3(k_1,k_2,k_3)
& \equiv
\sqrt{2}(k_1k_2)^2 \bigg[ -3 \bigg( \frac{1}{K^2} + \frac{2k_3}{K^3} \bigg)
+ \bigg( \frac{1}{K_1^2} + \frac{2k_3}{K_1^3} \bigg)
+ \bigg( \frac{1}{K_2^2} + \frac{2k_3}{K_2^3} \bigg) 
+ \bigg( \frac{1}{K_3^2} - \frac{2k_3}{K_3^3} \bigg) \bigg]
\, ,
\\
\label{eq:I4}
\calI_4(k_1,k_2,k_3)
& \equiv 
\frac{k_3^2-k_1^2-k_2^2}{\sqrt{2}} \bigg\{ 
3 \bigg( \frac{P}{K^2} + \frac{2k_1k_2k_3}{K^3} \bigg)
+ \bigg( \frac{P_1}{K_1^2} + \frac{2k_1k_2k_3}{K_1^3} \bigg)
\nonumber\\
&
+ \bigg( \frac{P_2}{K_2^2} + \frac{2k_1k_2k_3}{K_2^3} \bigg)
+ \bigg( \frac{P_3}{K_3^2} + \frac{2k_1k_2k_3}{K_3^3} \bigg) 
+ \log \bigg( \frac{K_1K_2K_3}{K^3} \bigg) \bigg\}
\, ,
\\
\label{eq:I5}
\calI_5(k_1,k_2,k_3)
& \equiv
\frac{\pi}{\sqrt{2}} \frac{k_1^2+k_2^2-k_3^2}{2} 
\, .
\end{align}
Note that unlike \cite{Bartolo:2017szm}, both helicity configurations, $s_1 = s_2$ and $s_1 \neq s_2$, contribute non-vanishing terms to the TTS bispectrum \eqref{eq:TTS}. Additionally, it is easy to see that in the BD limit, $\alpha \rightarrow 0$, we obtain $\mathcal{A}_1 = 0$ and $\mathcal{A}_2 = 2$. This recovers the result from \cite{Bartolo:2017szm} 
which should be proportional to $\mathcal{S}_5 \propto \pmb{k}_1\cdot\pmb{k}_2$  with $s_1 =s_2$. Meanwhile, the BD limit for opposite helicities 
is generally non-vanishing, which is different from the conclusion in \cite{Bartolo:2017szm}. 

We also notice that the contribution to the shapes in $\mathcal{I}_4$ contains a logarithmic scale dependence that diverges at the limits of certain shapes. Practically, this is not surprising. As explained at the start of this section,  both $H$ and $\mcs$ are assumed to be constants during slow-roll. These assumptions influence the precise form (i.e. explicit time-dependence) of the interaction Hamiltonian, which can, in certain cases, lead to artificial contributions when we take the limit $\eta \rightarrow 0$ in the in-in integrals. Consequently, this can generate a residual logarithmic divergence as it can be seen in \eqref{eq:I4}, which generally should not be taken seriously. 



\section{Analysis of the parity-violating cross-bispectra}
\label{sec:analysis}

%

In this section, we discuss in more detail the amplitudes and shapes of the cross-bispectra that we obtained in the previous section.

Before we proceed, a clarification is in order. The $\alpha$-vacuum formulation used here assumes scale-invariant initial conditions for inflation. Although this assumption is broken by the slow-roll expansion necessary to realize inflation, we expect its effects to manifest on almost all scales.
Furthermore, these initial conditions involve excited states, which can enhance the resulting amplitude and shape of the bispectra.
In a more realistic scenario, we would limit the enhancement to a finite range of scales. It will indeed be important to develop a comprehensive model and study the detectability of such a signal.
While this endeavour is very interesting, it lies beyond the scope of the present work. Our aim here is to demonstrate the potential for enhancing the parity-violating cross-bispectra through a simple modification of the initial conditions of inflation. 

\subsection{SST parity-violating cross-bispectra}


Similar to the standard definition, from \eqref{eq:Bpvsst} we introduce the non-linear parameter $\fnl$ as the bispectrum divided by the product of scalar and tensor power spectra:
\begin{equation}
\fnl^{s_3}(k_1,k_2,k_3) \equiv \frac{B^{s_3}(k_1,k_2,k_3)}{P_\calR P_\gamma} \, ,
\end{equation}
where $P_\calR \equiv 2\pi^2\calP_\calR/k^3$ and $P_\gamma \equiv \pi^2\calP_\gamma/k^3$. 
Looking at the shapes $\calS(k_1,k_2,k_3)$,  we can see that the dominant NG is due to sub-horizon effects with $k_1+k_2\approx k_3$ and $k_1 \sim k_2$ -- the flattened triangle configuration.  We expect the overall enhancement of the SST bispectrum will be determined by the suppression factor $H/\mcs$ to avoid the Chern-Simons instability and the amplitude function $\calA_1$. That is, focusing on the amplitude of $\fnl$ for general shapes, we can approximate it as
\begin{equation}
\fnl \sim \frac{H}{\mcs} \calA_1(\alpha,\psi) \, .
\end{equation}
%
%
Especially, as $\psi$ approaches $\pi$, the amplitude function $\calA_1$ is approximated as
\begin{equation}
\label{eq:A1-expansion}
\calA_1(\alpha,\psi\to\pi) 
\approx 
-\frac{\cosh(2\alpha)+\sinh(2\alpha)}{\big[\cosh(2\alpha)-\sinh(2\alpha)\big]^2} \sinh(2\alpha) (\psi-\pi)
\sim
-\frac{e^{8\alpha}}{2}(\psi-\pi) \, ,
\end{equation} 
so that $\fnl$ can be exponentially enhanced in all configurations. It is important to note that this is a relative enhancement, as the power spectrum is exponentially suppressed for $\psi\sim\pi$. Therefore, even if the tensor power spectrum remains unobservable, the cross-bispectrum might still be detectable.

%
%

\subsection{TTS parity-violating cross-bispectra}
\label{sec:TTS}

Next, we discuss the TTS cross-bispectra. Again, focusing on the $\fnl$ amplitude rather than the detailed shape dependence, we can approximate the contribution to the $\fnl$ due to the non-BD initial conditions to be proportional to $\mathcal{A}_1$ (for $\calS_1$, $\calS_2$, $\calS_3$ and $\calS_4$) or $\mathcal{A}_2$ (for $\calS_5$). As the phase $\psi$ approaches $\pi$, the amplitude of $\fnl$ can be exponentially enhanced, as $\calA_1$ in that limit is given by \eqref{eq:A1-expansion} and $\calA_2$ is approximated as
\begin{align}
\label{eq:A2-expansion}
\calA_2(\alpha,\psi\to\pi) 
& \approx 
\frac{3\sinh(4\alpha)+3\cosh(4\alpha)-1}{\big[\cosh(2\alpha)-\sinh(2\alpha)\big]^2}
\sim
3e^{8\alpha}
\, .
\end{align}
%
%
%
%
%
This effect can dominate for a large squeezing parameter $\alpha$. It was further argued in \cite{Kanno:2022mkx} that such a significant deviation from the BD vacuum  could  potentially be realized if we allow for a much lower value for the energy scale of inflation, e.g. hilltop inflation with $V_0^{1/4} \ll \mpl$ and in general, small field models with $\epsilon \ll 1$ and $H^2/\mpl^2 = V_0/\mpl^4$. As expected, this scenario would completely evade the Chern-Simons instability, as next-to-leading-order corrections to gravity are expected to experience a quadratic suppression of the form $(H/\mpl)^2$. This supression can be offset by a large squeezing parameter $\alpha$, where exponential enhancement is likely to dominate. This holds true for the conservative case where the scale for new physics for gravity is taken to be the Planck scale, though this is not necessarily required.  
Allowing for a much smaller cutoff scale $\Lambda$ for gravity (e.g., the string scale), could  easily realize an amplitude enhancement for the bispectrum. For instance, assuming an inflationary scenario  with an energy scale of $E=10^8$ GeV, we find that $H \sim E^2/\mpl \sim 10^{-3}$ GeV. The condition for the $\alpha$-vacuum scenario to be significant is  $H\ll \Lambda \ll E$. As one can see, $\Lambda$ in the range $\Lambda \in (10^5, 10^7)$ GeV would safely satisfy this condition.


%


We are now in the position to examine the bispectrum amplitude and corresponding shapes.  
Examining \eqref{eq:shapes}, \eqref{eq:I1}, \eqref{eq:I2}, \eqref{eq:I3}, \eqref{eq:I4} and \eqref{eq:I5}, we see that the shapes $\calS_1$ and $\calS_2$ scale as $1/K_i^4$. Thus, the dominant NG contributions are, for $s_1=s_2$, in the elongated ($k_1+k_2=k_3$) and flattened ($k_1=k_2=k_3/2$) configurations. But in this case because of the overall helicity factor, the cross-bispectrum with two left-circular polarization modes and that with two right-circular polarization modes have the opposite sign: 
\begin{equation}
B^{(++)}(k_1,k_2,k_3) = -B^{(--)}(k_1,k_2,k_3) \, .    
\end{equation}
This clearly exhibits the parity-violating signal due to the Chern-Simons term. Although the amplitude for both of these shapes is proportional to $f_\phi$ and is thus, as discussed in the previous section, suppressed by the Chern-Simons mass scale, this suppression can be offset by the relative enhancement given by \eqref{eq:A1-expansion}. 
Notice that there is also a subdominant contribution in the squeezed configuration, scaling as $k_3^{-3}$, where the long-wavelength scalar mode (i.e. $k_3 \rightarrow 0$) modulates the background. Due to its subleading nature, this contribution is less amplified, making the characteristic peak of the triangle in the elongated and flattened configurations, as discussed above, clearly distinguishable.
Further, we notice that the shapes $\calS_3$ and $\calS_4$ scale as $1/K_i^3$ and thus also peak in the same configurations as $\calS_1$ and $\calS_2$, but less dominant. Nevertheless, their contributions to the amplitude can be significant and may even dominate, as they are not suppressed by the Chern-Simons mass scale. In reality, it would be challenging to distinguish the difference in the slope of $1/K_i^n$ for various integer values of $n$ due to the finite resolution of CMB experiments.

\begin{figure}
 \begin{center}
 \begin{subfigure}[t]{0.45\textwidth}	
  \includegraphics[width=\linewidth]{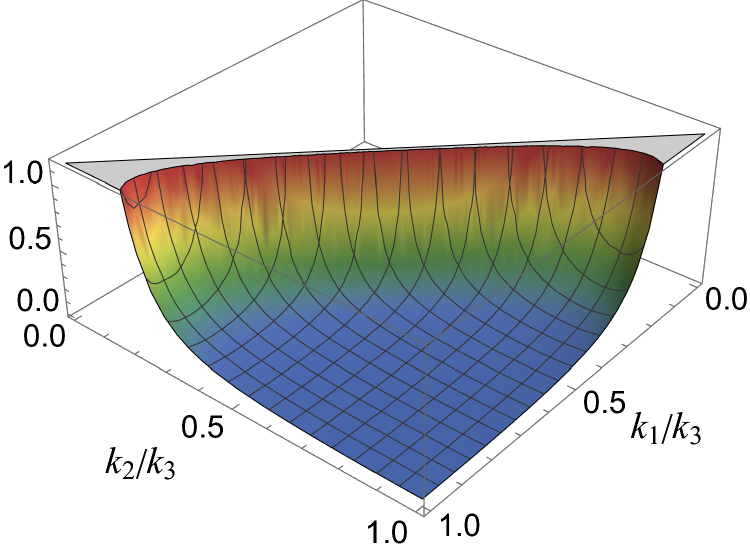}
  \caption{$\left[\calS_1^{(++)}+\calS_2^{(++)}\right]/10^4$}
 \end{subfigure}
 \qquad
 \begin{subfigure}[t]{0.45\textwidth}
  \includegraphics[width=\linewidth]{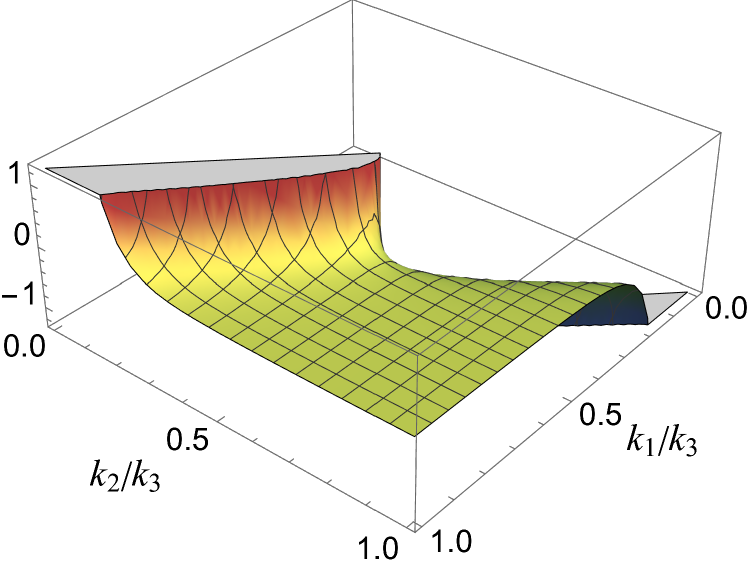}
  \caption{$\left[\calS_1^{(+-)}+\calS_2^{(+-)}\right]/(5\times10^3)$}
 \end{subfigure}
 \end{center}
 \caption{Shapes of the TTS cross-bispectrum with (left) $s_1=s_2=1$ and (right) $s_1=1$ and $s_2=-1$.}
 \label{fig:tts}
\end{figure}

Finally, in cases where the helicities differ (i.e. $s_1\neq s_2$), most discussions above for $s_1=s_2$ still apply. However,  due to the overall factor $k_1-k_2$ in \eqref{eq:shapes}, the bispectrum vanishes exactly at the flattened configuration and flips the overall sign across the line $k_1=k_2$.
In Figure~\ref{fig:tts} we show the shape functions $\calS_1+\calS_2$.

These parity-violating signals could be observed by measuring the BBT or BBE angular bispectra. 
\cite{Bartolo:2018elp} provides a Fisher matrix forecast on the detectability of a 
parameter $\Pi$ called {\it parity-breaking coefficient} using 
various 
angular bispectra of the CMB. It was showed that the minimum detectable value of $\Pi$ could be $\mathcal{O}(10^6)$, which can be achieved by allowing for a time-varying Chern-Simons mass. Unfortunately, a direct comparison with this work would be challenging because the Fisher matrix is typically scale-dependent and our predictions vary significantly compare to theirs for $\alpha \neq 0$. However, assuming that NG can be detected in the future, a Fisher forecast can help us develop a sense of how future experiments will react to various observables. That said, as this involves integrals over momenta, 
the pole structure of the bispectrum in the flattened configuration can present challenges. In this case, the ultraviolet divergence, as discussed in Introduction, must be specifically addressed.




 Finally, the divergent shapes discussed here were previously analyzed in \cite{Fergusson:2010dm}, in which they set WMAP-CMB constraints. The modal estimator, as detailed in \cite{Sohn:2023fte}, expands the bispectrum shape using mode functions but struggles with divergent shapes. To address this, a cutoff is introduced to limit the degree of flattening  of the shape (e.g. $K_{i} > k_c$ for $k_c \ll 1$) and the primordial shape is smoothed. Although this regularization introduces some uncertainty, it provides approximate constraints that capture the main characteristics of the shape. It would be interesting to perform this type of analysis for the parity-violating cross-bispectra in future work.


\section{Conclusions}
\label{sec:conc}

In this work, we computed the cross-bispectra for Chern-Simons gravity with non-BD initial conditions.
We identified a new shape for the parity-violating cross-bispectra, whose amplitude could experience an exponential enhancement in the flattened (folded) configuration, along with a significant contribution in the elongated configuration. 
While we also observe contributions in the squeezed configurations, these remain relatively subdominant. However, the relative enhancement described in \eqref{eq:A1-expansion} and \eqref{eq:A2-expansion} can lead to a significant improvement in the bispectrum amplitude compared to the case with  the BD initial conditions, potentially making the signal detectable in future surveys.

Given our choice to work with scale-invariant initial conditions, these effects manifest across nearly all scales. This is sufficient for  demonstrating the effect of excited initial states on the parity-violating three-point statistics, which is the focus of this work. 
In a more realistic scenario this effect could be restricted to a finite range of scales, which may even allow for further enhancement of the $\fnl$ amplitude beyond the scales relevant to the CMB observations, potentially entering the detection range of interferometers, such as LISA or PTA. We hope to revisit these questions in future work. 


\subsection*{Acknowledgments}

We would like to thank Shingo Akama, Xingang Chen, Kendrick Smith, Andrew Tolley and Yi Wang for useful discussions, which contributed to the improvement of this work. 
PC and JG are supported in part by the Mid-Career Research Program (2019R1A2C2085023 and RS-2024-00336507) through the National Research Foundation of Korea Research Grants. JG also acknowledges the Ewha Womans University Research Grant of 2024 (1- 2024-0651-001-1).
WL is supported by the National Research Foundation of Korea (2021R1C1C1008622).
MM is supported Kavli IPMU which was established by the World Premier International Research Center Initiative (WPI), MEXT, Japan. 
MS is supported by JSPS KAKENHI Grants 20H05853 and 24K00624.
PC is thankful to the Korea Institute for Advanced Science for its hospitality.
JG is grateful to the Asia Pacific Center for Theoretical Physics for its hospitality while this work was in progress.
MM is also grateful for the hospitality of Perimeter Institute where part of this work was carried out. Her visit to Perimeter Institute was supported by a grant from the Simons Foundation (1034867, Dittrich).
Research at Perimeter Institute is supported in part by the Government of Canada through the Department of Innovation, Science and Economic Development and by the Province of Ontario through the Ministry of Colleges and Universities.

\bibliography{bibimixed} 
\bibliographystyle{utphys}

\end{document}